# Performance monitoring for multicore embedded computing systems on FPGAs

Lesley Shannon*, Eric Matthews*, Nicholas Doyle*, Alexandra Fedorova[†]
* School of Engineering Science
Simon Fraser University, Burnaby, Canada
{lshannon, ematthew, ndoyle}@ensc.sfu.ca

[†] Department of Electrical and Computer Engineering
University of British Columbia, Vancouver, Canada
sasha@ece.ubc.ca

*Abstract*—When designing modern embedded computing systems, most software programmers choose to use multicore processors, possibly in combination with general-purpose graphics processing units (GPGPUs) and/or hardware accelerators. They also often use an embedded Linux O/S and run multi-application workloads that may even be multi-threaded. Modern FPGAs are large enough to combine multicore hard/soft processors with multiple hardware accelerators as custom compute units, enabling entire embedded compute systems to be implemented on a single FPGA. Furthermore, the large FPGA vendors also support embedded Linux kernels for both their soft and embedded processors. When combined with high-level synthesis to generate hardware accelerators using a C-to-gates flows, the necessary primitives for a framework that can enable software designers to use FPGAs as their custom compute platform now exist. However, in order to ensure that computing resources are integrated and shared effectively, software developers need to be able to monitor and debug the runtime performance of the applications in their workload. This paper describes ABACUS, a performance-monitoring framework that can be used to debug the execution behaviours and interactions of multi-application workloads on multicore systems. We also discuss how this framework is extensible for use with hardware accelerators in heterogeneous systems.

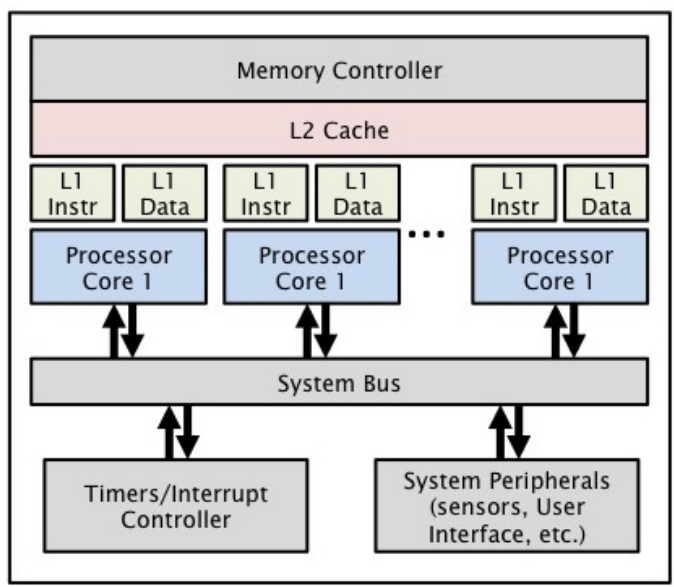

Fig. 1. Typical Multi-Core System Architecture

## I. INTRODUCTION

The complexity of embedded systems has increased dramatically in the past ten years. Software programmers now commonly design for embedded platforms with multicore processors running operating systems for multi-application workloads that are sometimes even multi-threaded. Recent developments in commercial Field Programmable Gate Array (FPGA) Computer Aided Design (CAD) flows and device architecture suggest that we are approaching the juncture where these software programmers may be able to shift to using FPGAs to implement these types of embedded systems and benefit from the inclusion of hardware accelerators.

Modern Field Programmable Gate Arrays (FPGAs) are large enough to implement hard and soft multicore architectures [1]–[3] in conjunction with custom hardware accelerators. FPGA vendors now also support High-Level Synthesis (HLS), which allows programs written in software languages (e.g. C) to be to be synthesized into actual hardware on a device. Furthermore, vendors provide Graphical User Interfaces (GUIs) that enable software programmers to describe their basic platform architecture and the inclusion of HLS-generated hardware accelerators; they also support embedded Linux kernels for both their hard and soft processors. As such, many of the necessary building blocks for software programmers to use FPGAs to build heterogeneous multicore computing systems already exist.

However, these building blocks unto themselves are insufficient for software programmers to use FPGAs to implement compute platforms at the level of abstraction with which they are comfortable. To enable a seamless transfer from more traditional multicore compute systems to heterogeneous multicore compute systems, software programmers require a complete design and development ecosystem that provides them with the *virtualization* and *visualization* to which they have become accustomed.

In this paper, we concentrate on a software developers need for visualization of an applications behaviour. Programmers need to be able to debug an application's functional behaviour as well as ascertain if (and potentially why) their current solution fails to meet necessary performance requirements. There is significant research ongoing as to how to provide software programmers the necessary functional debug support to develop software that executes as complete custom hardware or as a processor plus one or more hardware accelerators [4]–[8]. The remainder of this paper focuses on performance debug and runtime monitoring.

In a previous paper, we briefly presented a hArdware-Based Analyzer for Characterizing User Software (ABACUS) [9]. It focused on the need for a configurable microarchitectural independent hardware unit that could be used for workload investigation on a single processor core and processor architecture research on FPGAs [9]. This paper describes how the latest work on the ABACUS framework makes it a key visualization component for software programmers to use FPGAs for computing. We explain how ABACUS can be used to provide runtime performance information allowing a programmer to understand why a workload's execution is behaving as it is and what may be the source of its failure to meet performance requirements. We discuss its updated architecture and interface to both the user and OS. We also outline how this framework is not only for single core processors, but also is easily extensible to multicore systems, heterogeneous systems, and functional debug.

The remainder of this paper is organized as follows. Section II motivates why this type of framework is important for software programmers, particularly in heterogeneous systems and then highlights the requirements for such a framework to be accessible to software developers. Section III talks about the challenges and opportunities for performance visualization for software programmers using FPGAs as computing platforms. Related work is summarized in Section IV. We discuss our ABACUS framework in Section V, describing the architecture and system software as well as examples of different types of monitoring units. In Section VI, we outline how ABACUS is designed to be extensible to multicore and heterogeneous systems and can be used for functional debug. Section VII concludes the paper and recommends potential future work.

## II. Software Programmer Requirements for Performance Debug

When systems have multi-application workloads, the desire to share a single, coherent memory architecture often results in side effects. An obvious example for a single processor core system is the time needed for context switches to clear the state from the previously executing application and restore the state of the next application to execute. However, consider the case of a coherent multicore system, as shown in Figure 1. In this case, since applications on different processors may need to access memory at any given time – sometimes at the same time – these side effects may become more pronounced, while at the same time less predictable. These delays could negatively impact execution time it may not be able to fetch, process and store/display data in a timely fashion. As such, while the software may be functionally correct (i.e. performing the correct operations), it will not return meaningful results and fail to perform as required.

These types of problems often only arise after a period of execution as they result from software interactions from the various tasks executing on the system. This makes them much easier to detect and understand on the actual execution platform as opposed to in a simulator. However, for this to occur, software programmers must be provided the necessary infrastructure to monitor their system at runtime on the actual platform. First, the ideal infrastructure for this type of monitoring does not require any annotations of the actual software executable. This is because this annotation results in additional execution overhead, resulting in additional side effects. In some cases, this may exacerbate the existing problems. However, in other cases, it may cause new problems that do not exist in the actual system at runtime – or worse, the act of observing the system may artificially "correct" the performance bug, making it now undetectable. This is akin to someone having a bug in their program and then compiling it with the debug flags enabled, which somehow corrects the bug, making it extremely hard to find.

Ideally, this type of performance infrastructure will also have minimal impacts on the memory hierarchy when data is being stored for offline analysis. Depending on the nature and volume of data being stored about an application, it may need to be stored off-chip in the system's main memory. Obviously, this may also result in side effects in program execution as it may potentially introduce new contention to the system's memory arbitration.

For this type of performance monitoring and debug infrastructure to be usable to software developers on an FPGA, they must be able to configure and obtain the data via software *through the on-chip OS* while the FPGA is still configured with their hardware system. Ideally, the system monitoring units should be able to be reconfigured and restarted without having to re-tune the hardware or re-download the hardware system design. Instead, assuming the monitoring units are included as part of the actual system design, programmers can simply alter and update their software, scheduling assignments, etc. to alter system performance – without having to incur costly hardware redesign/re-synthesis time unless absolutely needed.

Additionally, the data obtained at runtime should be sorted and stored per process, if not per thread, so that the developer can track which specific applications (threads) are being penalized by the current resource/scheduling allocation so that this can be corrected more quickly. This type of data becomes even more crucial in heterogeneous systems where software designers may have a hardware accelerator that can be shared amongst a family of applications (e.g. a Discrete Cosine Transform for image processing applications). If both a software and hardware version of the accelerator exist, only a subset of the applications may actually require use of the hardware version to meet timing requirements.

## III. Challenges and Opportunities for Performance Visualization on FPGAs

Software programmers choosing to implement their designs on FPGA-based processor systems face some unique challenges; however, they are also provided with some unique opportunities. High-performance processors have long supported hardware counters for performance monitoring, and provide well-developed APIs to use these counters. Soft FPGA processors do not include hardware counters, however, the latest generation of FPGAs include embedded hard processors that do have hardware counters [1]. Unfortunately, hardware counters are often limited in number ($<< 100$), bit-width, and functionality, and can only be accessed when the processor is not executing application software.

In an FPGA-based system, however, it is possible to build an independent performance monitoring framework using some of the reconfigurable logic. This provides the opportunity for microarchitectural-independent data. Depending on the nature of the soft or hard processor being monitored, the user may be limited as to what signals can be monitored, but snooping a combination of the debug, interrupt, and memory access signals will provide software designers with much of the same information to which they have become accustomed.

In fact, by using the FPGA reconfigurable logic to build monitoring infrastructure, it is completely possible to build new performance monitoring units that are not specifically available. Using this approach, software designers can monitor considerably more complex behaviours than simple counters. Entire Block RAMs (BRAMs) can be allocated to store information such as: histograms of memory access patterns, complete data traces of parts of the programs, stall times and memory latencies, etc. Finally, in an FPGA-based system, the performance monitoring framework need not be integrated into the processor architecture. This allows it to altered/read before, during, or after program execution *and* it does not require any annotation of the software being monitored, reducing the chance for execution side effects.

## IV. RELATED WORK

At the accelerator level, there has been some work for visualizing and debugging HLS-generated accelerators. If software designers write their system description in C and then use HLS to generate the circuit, then waveform debugging is not as useful. Instead, a GNU-style debugger enables the programmers to visualize their solution in terms of variables and functions they had originally written as opposed to signals and circuits (and waveforms) created by the CAD flow [4]–[8]. Commercial vendors have also recognized the value of supporting on chip debugging of circuits [10], [11] and HLS designs [12].

Performance Monitoring for multicore systems is an active area of research [13]–[16]. Commercial vendors of multicore systems, such as AMD and Intel, also support profiling frameworks that use the hardware counters embedded in their System architecture [17], [18]. Additionally, previous researchers have also realized the value of using the FPGAs reconfigurable fabric to create additional instrumentation and monitoring circuitry to profile the system in operation, although this has generally been aimed at single core processors [9], [15], [19].

Xilinx has combined these two concepts to provide their SDSoC environment, which uses the embedded hardware counters in the ARM processor in conjunction with performance monitors instantiated in the reconfigurable fabric for monitoring performance on the bus [20]. This is the closest work to our ABACUS framework. However, unlike SDSoC, our performance monitoring framework does not require software to collect its data. Instead, it acts as a completely independent unit, with DMA support, able to write its data back to main memory, to reduce its impact on the actual system. It can also be used to generate interrupts to halt the CPU when specific situations are detected. Unlike SDSoC, ABACUS also has the necessary software to be used in conjunction with the system executing on the FPGA, communicating as needed with its OS. By reserving an OS page in the FPGA-based platform, when the data is uploaded to the OS page from ABACUS, the programmer or the OS can use and aggregate the data to make appropriate design and/or scheduling decisions. A final important feature to our system that is different than SDSoC is its extensibility. It is designed to enable users to create and/or select performance monitoring units from a library of units so that the performance monitoring infrastructure is best suited to the current application set of a particular workload. This extensibility is key to enabling it to support performance monitoring of heterogeneous systems.

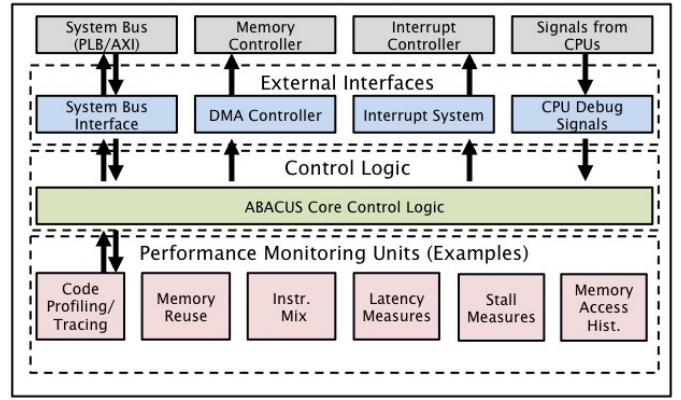

Fig. 2. ABACUS Architecture

## V. OUR ABACUS FRAMEWORK

Figure 2 illustrates the basic architecture of our improved ABACUS performance-monitoring unit. Similar to the original design in [9], we still maintain three basic system modules: External Interfaces, Control Logic, and the Performance monitoring units. The current design, however, supports a more complex external interface and more complex Performance Monitoring Units. In the original design, ABACUS was connected to the System Bus and snooped the desired signals from a CPU. Our new version is capable of snooping signals from multiple CPUs and associating the recorded data with a specific process or thread. It also supports DMA and has a corresponding device driver that can be included in the OS kernel. This enables the OS and the user to communicate directly to the device. ABACUS is assigned a specific address range on the system bus and is memory mapped by the OS. This enables status and configuration registers to be read or written through pointer referencing/dereferencing, to enable configuration settings such as reset, enable, and disable for specified processes or physical address ranges. The ABACUS device driver allows access through the ioctl function for reading/writing single registers as well as via mmap to access the full address space of ABACUS. The drivers enable the allocation of a page of kernel memory, enabling ABACUS to use DMA to copy its collected data into software space while the system is running.

The control level now supports the ability to record parallel data for a specific measurement being run on a multithreaded program executing on multiple processor cores concurrently. It also enables the user to time stamp when specific data was recorded (e.g. memory accesses in a data memory trace). It allows users to trigger ABACUS performance monitoring to start and stop based on various conditions, including the specific number of clock cycles or after a specific memory address access. For example, ABACUS can be triggered to collect user-specified data every time a specific instruction, function, or application executes. It is also able to send an interrupt to the OS to indicate that a specific situation has been detected based on user/OS configuration.

This latest design of ABACUS also includes more complex profiling units to demonstrate the true power and potential of an independent performance monitoring infrastructure not reliant on only hardware counters. For example, we have created a data memory access histogram unit to identify which regions of memory are accessed most frequently. When used in conjunction with our data memory access memory trace unit, this enables programmers to see which data regions are most frequently accessed and in which patterns, potentially facilitating a reorganization of data that reduces cache misses. We have created a memory latency unit to store a histogram of how many cycles it takes for each memory access to be completed. Another complimentary performance monitoring unit is our stall unit that measures the number of clock cycles a processor is stalled, waiting for the completion of an instructions.

This is just a subset of the potential units that can be designed and included as part of ABACUS. However, the key point is that it is easy for the user to select which of these units they wish to instantiate as part of their platform and configure the different parameters associated with each unit. It is also easy for the user to include new units as they are developed. It is possible to boot ABACUS with a set configuration, to start running once the system is powered up. However, the programmer or the OS is free to reconfigure ABACUS at runtime as desired, without having to reboot or re-download the system. Each of the individual performance monitoring units can then be configured independently and activated for any desired subset of the processor cores in the system.

## VI. EXTENDING THE ABACUS FRAMEWORK

We are currently extending the ABACUS framework in at least two directions:

**Multicore Scheduling**: Figure 3 illustrates how ABACUS has been integrated into our multicore PolyBlaze system [3]. Note that the *CPU Debug Signals* illustrated in Figure 2 can be used to monitor *any* signal in the processing system, even those internal to the processor, and their actual connections to the processor are dictated by the types of monitoring units that the user chooses to instantiate. As such, these connections have been excluded Figure 3, with the understanding that ABACUS can connect to *any* signal in a system that is deemed appropriate by the designer.

By including some of the processing units described in the previous section, we have been able to analyze multi-threaded workload execution across the processor cores by aggregating the results of an application's thread execution across the various processors. The key extension we have not yet completed for homogeneous multicore systems is how to use this data to guide the OS scheduler to potentially improve the platforms execution and power efficiency. Although the ABACUS performance monitoring data is currently accessible by the OS, it is not being used to influence scheduling decisions.

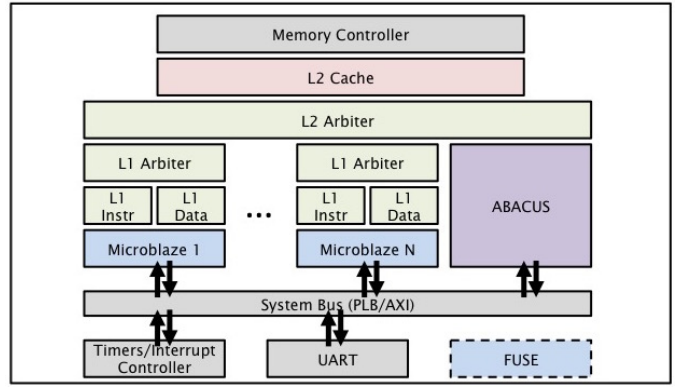

Fig. 3. Integration of ABACUS with PolyBlaze

**Heterogeneous Multicore Systems**: The ABACUS hardware framework and software infrastructure are already suitable for incorporation in a heterogeneous multicore system, as shown in Figure 4. The two components currently missing are specific performance monitoring units designed for hardware accelerators and the ability to provide ABACUS' runtime data to the OS to improve the sharing of hardware accelerators. Our current plan is to feed this data into the scheduler in FUSE (shown in Figure 3 in the dashed box), our Front-end USEr API (FUSE) for abstracting hardware accelerators in multicore systems [21]. The API could then use the data to share the hardware accelerators more efficiently and better meet the systems hardware performance requirements.

## VII. CONCLUSIONS AND FUTURE WORK

Much of the current research towards making FPGAs accessible platforms for software programmers focuses on HLS and bare metal programming (i.e. no OS) on single processor core systems combined with hardware accelerators. We believe that given the current prevalence of embedded multicore compute systems and the use of operating systems, it is important to consider this next step of development. As FPGA devices increase in size and complexity, with embedded multicore processors supporting operating systems, they become an obvious next generation technology choice for embedded computing systems to facilitate heterogeneity. Providing software programmers with the necessary hardware infrastructure to design these types of compute systems is insufficient to persuade them to adopt these platforms. Instead, we must provide them with the complete design ecosystem to which they have become accustomed, including OS support, debugging and performance monitoring. This paper focused on discussing the needs and opportunities for software designers to create multicore heterogeneous systems. We highlighted how software programmers could use our ABACUS framework in symmetric multicore and heterogeneous multicore systems to better understand the execution behaviour of their workloads to make better scheduling and design decisions.

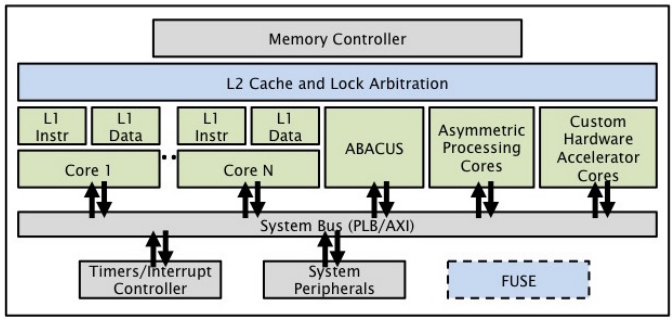

Fig. 4. Integration of ABACUS in a Heterogeneous System

In the future, we will be designing experiments to test ABACUS in heterogeneous compute environments. This will include integrating ABACUS support with a heterogeneous multicore platform virtualization API, such as FUSE [21] and then assessing what different types of performance monitoring units might be most appropriate. Based on this assessment, we hope to generate a basic framework for the key performance monitoring units so that they can be automatically included. The objective is to minimize the number of performance monitoring units software designers would need to generate for their individual systems to further facilitate their design process on a heterogeneous multicore compute platform.